\PassOptionsToPackage{capitalize}{cleveref} % Pass some extra options to `cleveref` if needed

\documentclass[acmsmall, nonacm]{acmart} %arxiv version

\usepackage{amsmath}
\usepackage{graphicx, booktabs, siunitx, float}
\usepackage[frozencache]{minted}
\usepackage{etoolbox}

\usepackage{version}
\excludeversion{podcout}

\includeversion{arxiv}
\excludeversion{cameraready}

\makeatletter
\AtBeginEnvironment{minted}{\dontdofcolorbox}
\def\dontdofcolorbox{\renewcommand\fcolorbox[4][]{##4}}
\makeatother

\newcommand{\kbp}{\mathbf{P}}

\newcommand{\decided}{\mathit{decided}}
\newcommand{\decides}{\mathit{decides}}
\newcommand{\decide}{\mathtt{decide}}

\newcommand{\init}{\mathit{init}}
\newcommand{\Agents}{\mathtt{Agt}}

\newcommand{\exchange}{\mathcal{E}}
\newcommand{\rimp}{\Rightarrow}
\newcommand{\dimp}{\Leftrightarrow}

\newcommand{\Nat}{\mathbb{N}}

\newcommand{\powerset}[1]{\mathcal{P}(#1)}
\newcommand{\Prop}{Prop}

\newcommand{\Estates}{S_e}
\newcommand{\I}{\mathcal{I}}

\newcommand{\R}{\mathcal{R}}

\newcommand{\messages}{\mathcal{M}}
\newcommand{\failures}{\mathcal{F}}
\newcommand{\adversary}{F}

\newcommand{\beln}{B^N} % belief, relative to being nonfaulty

\newcommand{\ebn}{E\hspace{-1pt}B_N} % everyone believes, amongst the nonfaulty
\newcommand{\cbn}{C\hspace{-1pt}B_N} % common belief amongst the nonfaulty
\newcommand{\ck}[1]{C\hspace{-1pt}K_{#1}}

\newcommand{\Values}{V}
\newcommand{\noop}{\mathtt{noop}}

\newcommand{\Time}{\mathit{time}}

\newcommand{\A}{\mathcal{A}}

\usepackage{xcolor}

\usepackage[ruled,vlined]{algorithm2e}
\newenvironment{program}[1][ht]
   { % Update algorithm name
    \begin{algorithm}[#1]
   }{\end{algorithm}}

\newcommand{\jdecided}{\mathit{jdecided}}
\newcommand{\deciding}{\mathit{deciding}}



\title{Model Checking and Synthesis for Optimal Use of Knowledge in
Consensus
Protocols}

\author{Kaya Alpturer}
\authornote{Alpturer was supported in part by AFOSR grant
 FA23862114029. Work done in part
while the author was studying at Cornell University.}
\affiliation{%
  \institution{Princeton University}%
  \city{Princeton, NJ}%
  \country{USA}%
}
\email{kalpturer@princeton.edu}
\orcid{0000-0003-4843-883X}

\author{Gerald Huang}
\authornote{Work of Huang conducted while at UNSW Sydney.}
\affiliation{
  \institution{University of Melbourne}
  \city{Melbourne, VIC}
  \country{Australia}
}
\email{gerald.huang@student.unimelb.edu.au}
\orcid{0009-0009-1014-2567}

\author{Ron van der Meyden}
\authornote{The Commonwealth of Australia (represented by the Defence Science and Technology Group) supported this research through a Defence Science Partnerships agreement.}
\affiliation{
  \institution{UNSW Sydney}
  \city{Sydney, NSW}
  \country{Australia}
}
\email{R.VanderMeyden@unsw.edu.au}
\orcid{0000-0002-9243-0571}

\acmYear{2025}\copyrightyear{2025}
\acmConference[PODC '25]{ACM Symposium on Principles of Distributed Computing}{June 16--20, 2025}{Huatulco, Mexico}
\acmBooktitle{ACM Symposium on Principles of Distributed Computing (PODC '25), June 16--20, 2025, Huatulco, Mexico}
\acmDOI{10.1145/3732772.3733552}
\acmISBN{979-8-4007-1885-4/25/06}

\begin{document}

\begin{abstract}
Logics of knowledge and knowledge-based programs provide a way to give abstract descriptions of solutions to problems in fault-tolerant distributed computing, and have been used to derive optimal protocols for these problems with respect to a variety of failure models. Generally, these results have involved complex pencil and paper analyses with respect to the theoretical ``full-information protocol" model of information exchange
between network nodes. It is equally of interest to be able to establish the optimality of protocols
using weaker, but more practical, models of information exchange, or else identify opportunities to improve their performance.
Over the last 20 years, automated verification and synthesis
tools for the logic of knowledge have been developed, such as the model checker MCK, that can be applied to this problem.
This paper concerns the application of MCK to automated analyses of this kind.
A number of  information-exchange models are considered, for
 Simultaneous and Eventual variants of Byzantine Agreement
under a range of failure types. MCK is used to automatically analyze these models.
The results demonstrate that it is possible to automatically identify optimization opportunities, and to automatically
synthesize optimal protocols. The paper provides performance measurements for the
automated analysis, establishing a benchmark for epistemic model checking and synthesis tools.
\end{abstract}

\maketitle

\section{Introduction}

Reasoning about multi-agent communicating systems can be subtle, particularly
in settings where agents and the communication media they use can be faulty.
Amongst the abstractions that have been developed to deal with the complexities in this
area are formal logics of knowledge \cite{HalpernM90,FHMVbook}, which provide a succinct way to express
properties of the states of information of agents operating in such environments.
It has been shown that for coordination goals, such as  consensus \cite{PSL},
formulas of the logic of knowledge can express the precise conditions in which agents can act to achieve these goals
\cite{DM90,MT88,HalpernMW01}. Incorporating these formulas into programs run by the agents yields what are called
\emph{knowledge-based programs} \cite{FHMVbook}. It has been shown that knowledge-based programs can exactly
characterize the behaviour of \emph{optimal} solutions for certain coordination tasks,
independently of the failure environment in which agents operate.
This has lead to the development of optimal protocols for these tasks in a range of
failure environments \cite{DM90,MT88,AHM23}.
The development of such results, however, requires non-trivial analysis to determine
concrete predicates of the local state of the agents that correspond to the situations
in which agents have the requisite knowledge in a given failure environment. A protocol that substitutes these equivalent
predicates for the knowledge formulas in a knowledge-based program is called an
\emph{implementation} of the knowledge-based program.

The application of this type of logical analysis of fault-tolerant distributed computing has been
conducted primarily in theoretical pencil and paper work. However, the applicability of logics of knowledge
has motivated work on the development of automated verification tools using such logics.
In particular, there has been work since the early 2000's to develop \emph{model checkers}
for the logic of knowledge \cite{GammieM04,LomuscioQR09}. Such a model checker takes as input a model of
a (typically, finite state) model of a protocol, and the environment in which it operates,
as well as specification in the logic of knowledge and time,
and automatically determines whether the protocol satisfies this specification in the given environment.
One of these model checkers, MCK, has moreover been extended to enable automated synthesis,
from knowledge-based programs, of distributed protocol implementations \cite{HMtark13,HM14tacas}.

The problems to which epistemic model checkers have been applied have  included
security protocols \cite{MeydenS04,Al-BatainehM10}, computer hardware protocols  \cite{BaukusM04}
and a range of applications in artificial intelligence \cite{HuangMM11}.
While fault-tolerant distributed computing played a key role in the emergence of the logic of knowledge
as an area of study in computer science, there has been comparatively little work on the application of
epistemic model checkers to this area. It is the objective of this paper to
begin a systematic study to address this lacuna.

We consider the application of the model checker MCK to the analysis of consensus protocols,
the most well-developed area for application of the logic of knowledge in fault-tolerant distributed computing.
Specifically, we focus on
variants of the problem of Byzantine Agreement \cite{PSL}.
In this problem, a
set of agents, some of which may be faulty, and each of which has a preferred value for some decision to be made, must come
to a consensus agreement amongst the non-faulty agents on one of the initial preferences.
In the Simultaneous variant of this problem (SBA) the agents must reach this decision simultaneously (in the same round of computation).
Solutions to the problem can be characterized
using a knowledge-based program in which agents agree upon a value when the non-faulty agents have
\emph{common knowledge} of the fact that a particular value is the initial preference of at least one agent \cite{DM90}.
Other knowledge-based programs express solutions to Eventual Byzantine Agreement (EBA), the variant where decisions are not required to be simultaneous \cite{HalpernMW01,AHM23}

The theoretical literature on the application of the logic of knowledge to Byzantine Agreement has
tended to focus on the analysis of ``full-information protocols'', in which agents repeatedly send their complete local state
to all other agents, and record in their local state all messages that they receive. This assumption
enables an analysis of the problem in which one develops protocols that are optimal in the sense that
agents make their decision at the earliest possible time, and such that no protocol (with any other
approach to information exchange) can systematically decide no later, and sometimes earlier. While the full-information
approach is useful for this theoretical purpose,  it is
not necessarily practical, since the full-information state of an agent grows at an exponential rate over time in Byzantine settings, and still quadratically when optimized in benign failure settings \cite{MT88}.
Indeed, in some circumstances, the optimal solutions must, moreover, perform intractable computations
at each step of the protocol
\cite{Moses09}.

Practical protocols, therefore, will operate with simpler state spaces and a lesser exchange of
information than a full-information protocol. Nevertheless, it remains of interest to determine
the optimality of such a protocol with respect to the reduced information exchanged. That is,
we can ask whether a protocol is optimal in its decision time, amongst protocols following the same
rule for information exchange. A formal definition for this notion of
optimality was given  for Eventual Byzantine Agreement by Alpturer, Halpern and van der Meyden \cite{AHM23},
who give a number of protocols for this problem that are optimal in this sense.
We can similarly define optimality for Simultaneous Byzantine Agreement with respect to a fixed information exchange.
It is shown in \cite{LIOPT} that the knowledge-based program of \cite{DM90} also characterizes SBA protocols that are optimal in this sense.

Reduced information-exchanges are also of benefit for model checking purposes, since complexity in the
state space of a model is one of the main factors impacting the computation run time
of a model checker.  Our contribution in this paper is to apply MCK in
an experiment on the feasibility
of
model checking the logic of knowledge as an approach to understanding Byzantine Agreement protocols
with limited information exchange. We do the following:
\begin{itemize}
\item
We develop formal models in the MCK scripting language
of a number of protocols for Byzantine Agreement, with limited information exchanges that have been proposed in
the literature, as well as the failure models in which those protocols operate.
We consider the following protocols:
\begin{itemize}
\item The FloodSet protocol of Lynch \cite{lynch}.

\item A protocol from \cite{CastanedaMRR17} in which agents additionally maintain a count of the number of agents
from which they received a message in the most recent round.

 \item A protocol from \cite{CastanedaMRR17} in which agents additionally maintain a count of the number of agents
 from which they received a message in the most recent round, as well as the previous
 value of that count.

 \item The concrete protocol of \cite{DM90} derived from an analysis of knowledge in the full information protocol for SBA in the crash failures model.

 \item  Protocols from \cite{AHM23} for the problem of Eventual Byzantine Agreement.
\end{itemize}
These models take as parameters a number of agents, an upper bound on the number of faulty agents,
and the number of possible values for the decision.
We
consider both the crash and the sending omissions failures models.

\item We apply MCK to determine automatically whether decisions are made in these protocols
at the earliest time that the agents achieve the required state of
knowledge.
We report results on how running times of the MCK model checking experiments scale
in the above-mentioned parameters.

\item In the course of these experiments, we identify a number of situations where protocols,
as proposed in the literature, are not optimal with respect to their chosen information exchange.
This means that there exist opportunities to optimize these protocols, by having agents make their
decisions earlier.

\item To better understand these optimization opportunities, and derive optimal protocols,
we use MCK to automatically synthesize implementations of the knowledge-based program for
Byzantine Agreement
with respect to the information exchanges and failure models considered.
We report results on how running times of the MCK model checking experiments scale
in the above-mentioned parameters. We informally describe the optimal protocols synthesized, in small instances,
but do not attempt to give a full characterization or a proof of optimality in the general case.
\end{itemize}

Taken together, the results of this paper show that, at least on small scale models,
it is feasible to investigate the question of optimality of a fault tolerant
distributed protocol by means of model checking using the logic of knowledge.
Moreover,  it is also possible to automatically synthesize optimal implementations of knowledge based programs,
for small scale models.

The structure of the paper is as follows. We begin in  Section~\ref{sec:knowledge} by laying out the semantic
framework for the logic of knowledge in which we work. Section \ref{sec:faults} describes the failure models we consider.
In Section~\ref{sec:spec} we give the formal specification for the
Byzantine Agreement (BA) problem.
The knowledge-based characterization of optimal solutions to the Simultaneous BA problem is described in Section~\ref{sec:simkbp}.
A knowledge-based characterization of optimal solutions to the Eventual BA problem is described in Section~\ref{sec:kbpeba}.
The model checker MCK is described in Section~\ref{sec:mck}.
Section~\ref{sec:protocols} describes a number of  protocols from the literature for solving SBA,
and the qualitative results that we obtain from our application of MCK model checking and synthesis to these protocols
using the characterization of Section~\ref{sec:simkbp}.
Section~\ref{sec:infexeba} presents a similar treatment for protocols for EBA.
In Section~\ref{sec:running} we describe the running times of our experiments.
Section~\ref{sec:related} describes related work and
Section~\ref{sec:concl}
concludes with a discussion of limitations of our results and possible future directions for research.
\begin{arxiv}
An appendix gives an example of the MCK synthesis input scripts used in our work,
and the result synthesized by MCK.
\end{arxiv}

\section{Semantic Model}\label{sec:knowledge}

To model distributed systems semantically, we first introduce the abstract interpreted systems model, with respect to which the
logic of knowledge and time has semantics. Some more specific instantiations of this model are described in the following section, in
which we describe how it can be used to model scenarios in which agents communicate by message passing and are subject to failures of various sorts.

\emph{Interpreted systems} \cite{FHMVbook} model multi-agent scenarios in which some
set  $\Agents$ of agents communicate and change their states over time.
An interpreted system is a  pair $\I = (\R, \pi)$, where $\R$ is a set of runs, describing how
the system evolves over time,  and
$\pi: \R\times \Nat \rightarrow \powerset{\Prop}$
is an interpretation function, that indicates which atomic propositions are
true at each {\em point} of the system, represented by a pair $(r,m)$
where $r\in \R$ is a run and $m\in \Nat$ is a natural number representing a time.
Each run $r\in \R$ is represented as a function
$r:\Nat \rightarrow \Estates\times \Pi_{i\in \Agents} L_i$,
where $\Estates$ is a set, representing the possible states of the environment in which the
agents operate, and where each $L_i$ is  a set,  representing the possible {\em local states} of agent $i$.
Given a run $r$, agent $i$ and time $m$, we write $r_i(m)$ for
the $i+1$-st component (in $L_i$) of $r(m)$,
and $r_e(m)$ for the first component (in $\Estates$).

Starting with a set of atomic propositions $\Prop$, we can build up a
logic, extending propositional logic, by introducing various types of
modal operators. Propositions are formulas, and, given a formula $\phi$, we have the following formulas:
$K_i \phi$ saying that agent $i$ knows that  $\phi$ holds, and
$\nu X(\phi(X))$ saying that the current situation is in the largest fixed point of an operator on
sets of points defined by the formula $\phi$.
(This is essentially the greatest fixed point operator from the linear-time mu-calculus \cite{Vardi88}, extended to
interpreted systems. It is required that propositional variable $X$ occur only in positive position for this operator to be meaningful.)

The semantics of the logic is given by a relation $\I,(r,m)\models \phi$,
where $\I$ is an intepreted system, $(r,m)$ is a point of $\I$, and $\phi$ is a formula,
defined inductively as
follows (we omit the obvious cases for the propositional operators):
\begin{itemize}
\item $\I,(r,m) \models p$ if   $p\in \pi(r,m)$, for $p \in \Prop$,

\item
$\I,(r,m)\models K_{i} \phi$ if  for all  points $(r',m')$ of $\I$  such that $r_i(m)  = r'_i(m')$  we have $\I,(r',m') \models \phi$.

%\item $\I,(r,m)\models \future\phi$ if   $\I,(r,m') \models \phi$ for all $m'\geq m$

\item If $\I = (\R, \pi)$, then
$\I,(r,m)\models \nu X(\phi(X))$ if   $(r,m')$
 is in the largest set of points $S$ of $\I$ such that $F(S) = S$, where
 $F(S)$ is defined to be the set of points $(r',m')$ such that  $(\R,\pi[X\mapsto S]),(r',m')\models \phi(X)$.
 Here $\pi[X \mapsto S]$ is the interpretation defined by $\pi[X \mapsto S](r,m) = (\pi(r,m)\setminus \{X\}) \cup \{X~|~(r,m) \in S\}$
 for each point $(r,m)$.

\end{itemize}

The intuition for the definition of the knowledge operator is that $r'_i(m) = r_i(m)$, says that agent $i$ considers it possible, when
in the actual situation $(r,m)$, that it is in situation $(r',m')$, since it is in the same local state there.
An agent then \emph{knows} $\phi$ is true in all the situations that the agent considers to be possible.

Using these operators, we can define a notion of common belief, that operates
with respect to an \emph{indexical set} $N$ of agents, which differs from
point to point in the system. That is, we assume that there is a function $N$ mapping each point of the system to a set of agents.
The semantics of the atomic formula $i\in N$ is given by
 $\I,(r,m) \models i \in N$ if   $i\in N(r,m)$.

An agent may not know whether it is in a set $N$. We can define a notion of belief, relative to the indexical set $N$, by $\beln_i \phi = K_i(i\in N \rimp \phi)$.
Using this, we define the notions of ``everyone in $N$ believes''
$\ebn \phi = \bigwedge_{i\in N} \beln_i \phi$.
Common belief, relative to an indexical set $N$, is defined by
$\cbn \phi = \ebn \phi \land \ebn^2 \phi \land \ldots$.
Equivalently, $\cbn\phi = \nu X( \ebn (X \land \phi)))$.
It is immediate from the fixpoint characterization that $\cbn \phi \equiv \ebn \cbn \phi$.
Provided it is valid that
$N \neq \emptyset$, we have that
$\ebn\phi \rimp \phi$ and
$\cbn\phi \rimp \phi$.

\section{Communication and Failure Models} \label{sec:faults}

For the problems we consider, it is convenient to consider a two-layer protocol model comprised of
an information exchange protocol $\exchange$ as the base layer, over which we run a decision protocol $P$.
Intuitively, the information exchange protocol $\exchange$ defines the agents' local states,
the messages they exchange, and how these states are updated.
On the other hand, the decision protocol $P$ dictates the actions of each agent in a given round.
In the context of the Byzantine Agreement problem,
an action will either be $\noop$, representing no action,
or $\decide_i(v)$ which represents a decision on a specific value $v$ by agent $i$.
(The details of the model are as presented in \cite{AHM23}. As we do not develop proofs in this paper, we give here just an informal presentation, and refer the reader interested in the details to that paper.)
Nonfaulty agents run both the information exchange and decision protocols correctly, but faulty agents may deviate from these protocols, in ways that depend on the failure model.
A further parameter is the failure model $\failures$.  Thus, our systems are denoted as $\I_{\exchange,\failures,P}$,  but we may elide the parameters when they are clear from
the context.

For the communications model we assume  message passing as the communications medium.
Each agent may send messages from a set $\messages$.
Our focus will be on \emph{synchronous} message passing, in which agents operate in a sequence of synchronized rounds.
In each round, each agent sends a set of messages to the other agents,
receives some of the messages from the other agents that were sent  in the same round, and updates its state depending on these events.
The information exchange protocol $\exchange$ describes the possible initial states of each agent
(which may include information such as the agent's preference for the
outcome of the consensus decision to be made), how it chooses the messages to be sent in each round (which may depend on its local state
and the action performed in the round),  and how it updates its state in response to its actions and the messages received in a round.

The failures model describes what failures  can occur.
Typically, a failures model comes with a parameter $t$ that indicates the maximum number agents that may be faulty. The possibility of solutions to consensus problems generally depends on $t$ in some way,
e.g., for Byzantine failures, solutions for consensus using deterministic protocols are only possible when $3t$ is less than the number of agents $n$ \cite{PSL}.
A \emph{failure pattern},  or \emph{adversary}, $\adversary$ expresses the failures that actually occur in
a particular run.
Each failure model is associated with a set of adversaries.
We consider the following failure models, parameterized by  an upper bound of $t$ on the number of faulty agents:
\begin{itemize}
\item {\tt Crash($t$)}: in this model, agents may fail by crashing. When an agent crashes in round $m$,
it sends an arbitrary subset of the messages it was supposed to send in round $m$. In later rounds, it sends no more messages.
\item {\tt Sending-Omissions($t$)}: a faulty agent may fail to send any message that it was supposed to send, but receives
all messages that have been sent to it.
\end{itemize}
(The knowledge-based programs we consider can also be applied to
other failure models, such as
receiving and general omissions failures.)

A decision protocol $P$
is a function from the agent's local state to its action in the next round.

To connect the above sketch to the semantic model of Section~\ref{sec:knowledge}, we describe how an information exchange protocol $\exchange$, a failure model $\failures$ and a
decision protocol $P$ determine an interpreted system $\I_{\exchange,\failures,P}$. In this system, states of the environment $\Estates$
record a failure pattern $\adversary$ from $\failures$.
An \emph{initial global state} consists of a failure pattern $\adversary$ for the environment state,
and for each agent $i$, an initial state of the information
exchange protocol in  $L_i$.  For each initial global
state, a run $r$ with that initial state is uniquely determined by the information exchange protocol $\exchange$,
the failure model $\failures$, and the decision protocol $P$. In each round of this run,
the decision protocol $P$ determines  what action each agent performs in the round, and
the information exchange protocol $\exchange$ determines what messages agents are required to  send.
The failure pattern $\adversary$ determines which of these messages are actually sent, and which are
actually received. As a result of the action taken and the messages received,
 each agent updates its local state as required by the information exchange protocol $\exchange$.
The set of runs of $\I_{\exchange,\failures,P}$ consists of all runs generated in this way from some initial global state.

\section{Specifications} \label{sec:spec}

We consider consensus problems in which each of the agents in the
system is required to make a decision on a value in some set $\Values$,
and the agents are required to ensure that they make the \emph{same} decision.
There exists a range of definitions of this problem in the literature.
We focus first on a version
where \emph{nonfaulty} agents are required to make their
decisions consistently and \emph{simultaneously}.
We later consider a version where they may decide at different times.

The  local states $L_i$ of agent $i$ contain the agent's initial preference $\init_i\in \Values$ for the decision to be made, as well as components
derived from the messages that the agent has received. (Details of these components and how the agent updates them
depend on the  specifics of the information exchange protocol. We give a number of examples below.)
Each agent $i$ has, at the start of a run, an initial value $\init_i \in \Values$ representing its preference for the
decision to be made.  The possible actions of agent $i$
are $\noop$ and $\decide_i(v)$ for each $v\in \Values$, representing agent $i$'s action of making the decision on value $v$.
We write $\decides_i(v)$ for the proposition stating the next action that agent $i$ performs according to its decision protocol is $\decide_i(v)$.

Some of the agents may be faulty; we write $N$ for the set of nonfaulty agents.
We note $N$ is \emph{indexical}, in the sense that different runs have different values for the set $N$. The set of
faulty agents, and how they fail, is given in the adversary $F$ of the run.
We consider the following requirements:
\begin{itemize}
\item {\bf Unique-Decision:} Each agent $i$ performs action $\decide_i(v)$ (for some $v$) at most once.
\item {\bf Simultaneous-Agreement(N):} If $i\in N$ and $\decides_i(v)$ then, at the same time, $\decides_j(v)$ for all $j\in N$.
\item {\bf Validity(N):} If $i\in N$ and $\decides_i(v)$ then $\init_j(v)$ for some agent $j$.
\end{itemize}

A protocol guaranteeing these properties is said to be a \emph{Simultaneous Byzantine Agreement} (SBA) Protocol.
(We remark that termination is not a requirement of this specification. There exist knowledge-based characterizations of
the problem where various termination and alternative agreement and validity properties are used, but we do
not delve into these alternatives in the present paper.
\begin{arxiv}
We do, however, investigate protocols that are guaranteed to terminate and verify termination
properties in our scripts as in Appendix~\ref{app:floodset}.
\end{arxiv}
\begin{cameraready}
We do, however, investigate protocols that are guaranteed to terminate and verify termination
properties in our scripts as in the full version of the paper \cite{full-version}.
\end{cameraready}
)

{\bf Optimality:} A concern in the literature has been to develop SBA protocols that are optimal in the sense that decisions are made as early as possible.
In order to compare two protocols, we need to be able to be able to compare runs of the two protocols.
This has typically been done in the literature by comparing protocols (with an arbitrary information exchange) with the full information protocol, and
showing that the full information protocol is optimal \cite{DM90,MT88,FHMVbook}.
We consider here a definition of optimality that compares protocols using the \emph{same} information exchange.
A definition of this kind was first given in \cite{AHM23} for Eventual Byzantine Agreement protocols.
We consider here a similar definition from \cite{LIOPT} for SBA protocols.

 We rely here on the fact that the
transitions are deterministic once deterministic protocols $P$ and $\exchange$ and an initial global state (containing an adversary $F$ that resolves all nondeterminism) have been identified.
Given an information exchange protocol $\exchange$ and a failure model $\failures$,
we say that a run $r$ in $\I_{\exchange,\failures,P}$ of a decision protocol $P$
\emph{corresponds} to a run $r'$  in $\I_{\exchange,\failures,P}$ of decision protocol $P'$,
if $r$ and $r'$ have the same initial global state.
This means that all agents start with the same initial preferences, and
face the same pattern of failures over the two runs.

Using this, we can define an order on decision protocols $P,P'$ with respect to an information exchange protocol $\exchange$ and failure model $\failures$.
Define $P \leq_{\exchange,\failures} P'$ if for all corresponding runs $r$ of $\I_{\exchange,\failures,P}$ and $r'$ of $\I_{\exchange,\failures,P'}$, and for all %\emph{non-faulty}
agents $i$, if  agent $i$ decides at time $t$ in $r$ then agent $i$ does not decide earlier than $t$ in run $r'$.

A decision protocol $P$
is \emph{optimum} for SBA relative to  $\exchange$ and $\failures$,
if it is an SBA protocol for $\exchange$ and $\failures$, and for all SBA protocols $P'$ for $\exchange$ and $\failures$, we have
$P \leq_{\exchange,\failures} P'$. That is, $P$ always makes decisions no later than any other SBA protocol for $\exchange$ and $\failures$.

A  decision protocol $P$ is \emph{optimal} for SBA relative to  $\exchange$ and $\failures$, if it is
an SBA protocol  for $\exchange$ and $\failures$, and  for all SBA decision
protocols $P' \leq_{\exchange,\failures} P$ we have $P \leq_{\exchange,\failures} P'$.
That is, there is no decision protocol for $\exchange$ and $\failures$  in which decisions are made no later than they are in $P$, and sometimes earlier.

\section{Knowledge-Based Program - Simultaneous Byzantine Agreement}

 \label{sec:simkbp}

For the problem of Simultaneous Byzantine Agreement, we work with
the knowledge-based program for SBA from \cite{MT88,FHMVbook}.
However, whereas these works focus on full information implementations,
we will also consider weaker models of the underlying information flow.

For a value $v\in \Values$, we write $\exists v$ for $\bigvee_{i\in \Agents} \init_i=v$.
The results of \cite{MT88,FHMVbook}
lead to the following knowledge-based decision program $\kbp$, for each agent $i$:
\begin{equation}
\begin{array}{l}
\mbox{do $\noop$ until}~\exists v\in V (\beln_i \cbn \exists v);\\
\mbox{let $v$ be the least value in $V$ for which}~\beln_i \cbn\exists v  \\
\decide_i(v)
\end{array} \label{simkbp}
\end{equation}
This program characterizes the conditions under which decisions are made
in a way that abstracts from the concrete details of how the environment operates, how information exchange protocol states are maintained,
and how agents compute what they know. (The program can be shown to be equivalent to a program in \cite{DM90}, for the crash failures model,  that
uses instead the knowledge conditions $K_i \ck{\A} \exists(v)$, where $\A$ is the set of agents that are \emph{active}, that is, have not yet crashed.)

Knowledge based programs are  not directly executable, but are more like specifications that need to be implemented by replacing the knowledge tests by concrete predicates of the agent's local states.
We refer to \cite{FHMVbook} (Chapter 7) for motivation and detail.
An implementation $P$ of
the knowledge-based program $\kbp$,
relative to an information exchange protocol $\exchange$ and a failure model $\failures$,
is obtained by substituting concrete predicates
of the local state of each agent $i$
for the knowledge conditions $\beln_i \cbn \exists v$,
that are equivalent to these knowledge conditions in the interpreted system $\I_{\exchange,\failures,P}$.

The knowledge-based program
$\kbp$
is shown to be correct in \cite{MT88} in the sense that if any
SBA protocol
exists in a full information exchange context, then
an implementation of
$\kbp$
solves this problem in that full information exchange context.
Moreover,  an implementation of this program is optimum in this context.
Concrete implementations using the full information exchange protocol
have been derived for crash failures \cite{DM90} and for omission failures \cite{MT88}.

In our work on model checking and synthesis, we model the knowledge-based program $\kbp$ in a variety of failure environments
and information exchange protocols. A correctness and optimality result can be established
\cite{LIOPT} that
generalizes that of \cite{DM90,MT88}.
\begin{podcout}
Whether we obtain an optimal or an optimum protocol as the implementation of the
knowledge-based program depends on the information exchange protocol $\exchange$.
We say that $\exchange$ \emph{does not transmit information about actions} if the messages
transmitted by an agent, and the state updates that the agent performs, do not depend on the action the agent performs in the round.
We say that $\exchange$ \emph{does not transmit information about decisions} if the messages
transmitted by an agent, and the state updates that the agent performs,
may depend on whether the agent performs $\noop$ or a $\decide_i(v)$ action, but are the
same for all actions $\decide_i(v)$ and $\decide_i(v')$ for $v,v'\in \Values$.
The following result from \cite{LIOPT} enables us to determine whether implementations of
$\kbp$ we find are optimal or optimum:
\begin{proposition}
Let $\exchange$ be an information exchange protocol and let $\failures$ be failure model, and
let protocol $P$ be an implementation of $\kbp$ relative to  $\exchange$ and $\failures$.
\begin{itemize}
\item If $\exchange$ does not transmit information about actions, then
$P$ is  an optimum SBA decision protocol relative to $\exchange$ and $\failures$.
\item If $\exchange$ does not transmit information about decisions, then
$P$ is an optimal SBA decision protocol relative to $\exchange$ and $\failures$.
\end{itemize}
\end{proposition}

\end{podcout}

\section{The Model Checker MCK} \label{sec:mck}

Model checking is an automated verification technology \cite{clarkebook} applicable to concurrent systems. Given, as  input, code for the
behaviour of a concurrent system and a formula in a modal logic,  a model checker will automatically check whether the formula is true of all runs of
the system. The modal logic used in model checking has traditionally been a form of temporal logic. \emph{Epistemic} model checkers \cite{GammieM04,LomuscioQR09}
extend this technology to specifications in a logic of knowledge and time.
MCK \cite{GammieM04}, the model checker we use in this paper, supports linear time and branching time temporal logic, knowledge operators,
as well as least and greatest fixpoint operators including the operator $\nu$ defined above.

Concurrent systems are described in MCK by describing an environment, declaring the agents in the system and the protocols run by each of the agents.
The environment is described by declaring variables for the environment states in $S_e$, specifying
(by means of a formula or code) which of these states are initial states, and giving nondeterministic code that computes the transitions
on the environment states. This code may take as input the actions performed by the agents in a round of the computation.
Since these actions are treated as being performed simultaneously, but transition code execution is treated as taking a single step of
computation, the MCK modelling language is in the class of \emph{Synchronous languages}.
All variable types in MCK (boolean, enumerated or finite numerical ranges) are finite; this implies that the model is finite state.

A description of an agent's protocol includes the binding of the protocol's input parameters to environment variables, a declaration of
additional local variables, and code that describes the action performed at each round of the computation of the system.
Actions may be a sequence of assignments to local and parameter variables or a signal sent to the environment as an
input to the environment transition code.

Some of the agent's variables may be declared to be \emph{observable}. The collection of values of the agent's observable variables
makes up the agent's \emph{observation}.
A number of different semantics for the knowledge operators
are supported in MCK, depending on the degree of recall that an agent has of its observations.
In the present paper, we apply the \emph{clock} semantics, which states that an agents's local state in $L_i$ is a
pair $(\Time,o)$, where $\Time\in \Nat$ is the current time, and $o$ is the observation that the agent makes at that time.
A formal description of how the MCK modelling language relates to the interpreted systems model is given in the
MCK manual \cite{mckManual}.

In addition to model checking, MCK now supports automated synthesis of implementations of knowledge based  programs.
In a synthesis input to MCK, agent protocols may declare boolean \emph{template variables}, which function as
placeholders for predicates over the agent's local variables, which will be automatically synthesized.
The template variables may be used in conditions within the agent's protocol code. For each template variable $x$, there is a \emph{requirement} that
relates $x$ to a formula of the logic of knowledge. This enables knowledge based programs such as the program $\kbp$ above to
be expressed in MCK.

Synthesis algorithms for a number of different semantics of knowledge have been developed \cite{HuangM13,HuangM14,HuangM16}. In the present work,
we apply the algorithm for the clock semantics of knowledge.
In this case, the requirement for agent $i$'s template variable $x$ (at the time that the template variable is
used in the agent's protocol), is of the form $x \dimp \phi$ where $\phi$ is a boolean combination of formulas of the form $K_i \phi$, which may not
contain temporal operators, but may contain knowledge operators and fixpoint operators.
The formulas $\beln_i\cbn \exists v$ satisfy these constraints, so we may write the knowledge based program for SBA in
MCK's input language.
\begin{arxiv}
An example of an MCK script providing a synthesis input for the SBA problem (for the FloodSet information exchange described below)
is given in Appendix~\ref{app:floodset}.
\end{arxiv}
\begin{cameraready}
An example of an MCK script providing a synthesis input for the SBA problem (for the FloodSet information exchange described below)
is given in the full version of this paper \cite{full-version}.
\end{cameraready}

In general, a knowledge based program may have zero, one or many implementations \cite{FHMVbook}.
Theoretical results of \cite{FHMVbook} imply that for the clock semantics, knowledge based programs, subject to the constrains imposed by MCK,
have a unique implementation. MCK's synthesis algorithms automatically compute the predicates on the agent's local states that,
when substituted for the template variables, yields the concrete protocol that implements the knowledge based program.
Both model checking and synthesis algorithms in MCK are implemented using Ordered Binary Decision Diagram techniques~\cite{BCDHM90}.

\section{Information Exchange Protocols for SBA} \label{sec:protocols}

We now describe the information exchange protocols that we have modelled in MCK, and with respect to which we have
conducted an evaluation of MCK for model checking and synthesis from knowledge based programs.
The present section describes information exchanges we have considered for the SBA problem.
Section~\ref{sec:infexeba}
considers information exchanges for EBA.
 In the present version of the paper, we report results for the crash and sending omissions failures model.
In the following, $n$ denotes the number of agents in a scenario, and
 $t$ denotes the maximum number of these agents that may crash during the running of the protocol (so $t\leq n$).
 It is well-known that in some runs, a decision cannot be made before round $t + 1$,
so protocols often defer a decision to that round.

As protocols are usually modelled, decisions in round $t+1$ are made after all the messages from that round have been received.
 In our modelling, to determine an agent's knowledge at the end of a round,
 we capture an agent's state at the end of round $m$ as the state at time $m$, so the decision
 would be made in round $t+2$,
  as a function of knowledge computed at time $t+1$.
 We are interested in determining the earliest possible decision time
 in each run, given the information being exchanged by the protocol.

\subsection{FloodSet Protocol}

The FloodSet protocol is described in Section 6.2.1 of Lynch's text {\em Distributed Algorithms} \cite{lynch}
(simplifying ideas from other protocols in the literature).
Each agent maintains a set of values that it has seen. Initially, each agent will only have seen its own initial value. In each round, each non-faulty agent broadcasts the set of all values that they have historically seen, and updates their set of values by adding all of the values in messages received in the current round. A decision is made on the lowest
value received by the end of round $t+1$.

In our modelling, the local state of agent $i$ has the form
$\langle w,\mathit{time}\rangle$, consisting of
an array $w:\Values \rightarrow \mathit{Bool}$ indicating which values have been seen,
and the current time $\mathit{time}$ (number of rounds executed).
Based on the presentation in \cite{lynch}
where the stopping condition is $\mathit{time} = t+1$,
one expects that the earliest time at which
the condition $\beln_i \cbn \exists v$ holds for some value $v$ is $t+1$.
Our model checking experiments automatically identify a situation in which this \emph{false} in some runs.
For example, it is not true in the case of $n=3$ and $t=2$.

We have conducted a theoretical analysis which shows that in the case $t\geq n-1$,
the common knowledge condition $\beln_i \cbn \exists v$ holds for some value $v$ already
at time $n-1$, so a decision can be made earlier. This results in a revised hypothesis for the
earliest time at which the condition $\beln_i \cbn \exists v$ holds with respect to this information exchange, namely
\begin{equation}
(t \geq n-1 \land \mathit{time} = n-1) \lor (t < n-1 \land \mathit{time} = t+1) \label{eq:floodcond}
\end{equation}
The model checking experiments we have conducted support this hypothesis:
it is reported as true in all the instances we  were able to check.%
\footnote{We have developed a
theoretical proof that this condition captures, for all $n$ and $t$, the
earliest time at which the common knowledge condition is satisfied for the FloodSet protocol
\cite{AMRW24}.
Although, to our knowledge, the result has not been stated in this precise
form in the literature, we note that a construction in \cite{DolevRS90}
shows that a protocol that is correct for up to $t' = n-2$ crashes
can  be transformed into one that is correct for up to $t=n$ crashes.
The effect of this construction is a protocol that stops at time $t'+1 = n-1$ in this case.}

When we apply MCK to automatically synthesize an implementation of the knowledge based program $\kbp$,
we find that it synthesizes a decision condition that is equivalent to condition~(\ref{eq:floodcond}) in
all the cases we were able to check.
\begin{arxiv}
(The appendix gives an example of an MCK synthesis
script and the result produced by the model checker.)
\end{arxiv}

\subsection{Counting the number of crashed agents}

A number of variants of the information exchange in the FloodSet protocol are
described by Casta\~neda et al. \cite{CastanedaMRR17}, in a study of a range of
early stopping conditions. The specification considered in that paper
is for \emph{Eventual} Byzantine Agreement (EBA) --- we consider the information exchanges instead for the
simultaneous  specification SBA,
in order to investigate if they also admit early stopping conditions in that case.
(One of the benefits of addressing this question by model checking and synthesis is that evidence for the answer can, in small instances, be obtained automatically, without the cost of the mental effort to construct a proof.)

One of these variants sends the same messages as in the FloodSet protocol, but
has each agents also keep a count of the number of agents that it knows have {\em actually} crashed.
Since a message is broadcast in each round, for a pair of agents $i$ and $j$,
if agent $i$ does not receive a message from agent $j$ in a round, then $i$ knows that $j$ has crashed.
Each agent maintains a variable $\mathit{count}$ that is updated in each round to be
 the number of messages received by the agent in the round.
 An agent is treated as sending itself a message in each round. An agent's local state in this model
 has the form $\langle w, \mathit{count},\mathit{time}\rangle$, where $w$ and  $\mathit{time}$ are
as in the FloodSet model.

 In view of the conclusions above about the FloodSet information exchange, a reasonable null hypothesis is
 that condition~(\ref{eq:floodcond}) captures the earliest time at which $\beln_i \cbn \exists v$ holds for some $v$,
 so that a decision can be made. Model checking shows this to be \emph{false}, indicating that
 that this information exchange protocol gives the agents extra information that allows an earlier stopping time.

Plainly,  if $\mathit{count} \leq 1$ (something that is possible only if $t\geq n-1$), then this implies that at most one message has been received by the agent. Since the agent {\em always} receives their own message, this implies that every other agent must have crashed and so, it is safe to make a decision at the current round. Correspondingly, when just one agent remains, there is common knowledge amongst the nonfaulty agents
reduces to that agent's own knowledge, and common knowledge of a value necessarily holds.
This gives an immediate early exit condition: $\mathit{count} \leq 1$.

For this model, both our model checking and synthesis experiments confirm that the earliest time at which the
common knowledge condition
for SBA
holds is when
\begin{equation}
\mathit{count} \leq 1 \lor (t \geq n-1 \land \mathit{time} = t) \lor (t < n-1 \land \mathit{time} = t+1)
\label{eq:condcount}
\end{equation}
In particular, even condition
$\mathit{count} \leq 2$
does not suffice to enable a decision unless the FloodSet
condition~(\ref{eq:floodcond}) holds.

Model checking and synthesis can give information only about small numbers $n$ of agents, but we have subsequently
validated this result for general $n$ using a theoretical analysis \cite{AMRW24}.

\subsection{Diff: Memory of Count}

A further protocol $P_{\mathit{diff}}$ considered by by Casta\~neda et al. \cite{CastanedaMRR17}, makes use of not just a count of the number of messages received,
but also remembers the value of this variable from the round before the last. It is shown that the difference between these values can be used to
give an earlier stopping condition for EBA than that obtained by using a count alone. We have also modelled this information exchange. The only modification
required to the model for the single count is to add another variable for the earlier round count and to assign the count value to this variable at the start of each round,
before determining the number of messages received in the current round.

While for the EBA problem, \cite{CastanedaMRR17} show that for the crash failures model,
the difference between the most recent count of messages received and the previous value of this variable
 allows early decisions to be made,
in our model checking experiments for this information exchange and failure model, we did not find a condition allowing a decision for the SBA problem
that is stronger than that for the version with just a single count variable, as in the previous subsection.
We have subsequently validated this result using a theoretical analysis for the general case of an arbitrary number of agents
\cite{AMRW24}.

\subsection{Dwork-Moses}

Finally, we have  modelled the protocol derived by Dwork and Moses \cite{DM90} as a result of an analysis of common knowledge in the
full-information protocol. (The protocol is presented in Fig.~2 of that paper.)   This protocol works for the set of decision values $\{0,1\}$,
In this model, we do not attempt to represent the full-information state; rather, we represent just the variables of the protocol of Dwork and Moses.
These consist of variables $F,NF,RF$, which are sets of agents, representing the set of agents known to be faulty, the set of agents
newly discovered by the agent to be faulty, and the set of faulty agents that the agent has heard about from  other agents.
Each agent also maintains a variable $\mathtt{exists0}$ representing whether it is aware that some agent has initial value 0.
In each round, the protocol broadcasts the pair $(NF,\mathtt{exists0})$. There is also an integer variable $\mathtt{current\_waste}$
that represents the agent's estimate of the number of failures that have been \emph{wasted} in the current run, where a failure is wasted if
it was not needed to delay a clean round.

Intuitively, the amount of waste can be used to determine that there has been a \emph{clean} round, meaning a round in
which no new failures were detected. After a clean round, all nonfaulty agents have received the same set of values,
so are guaranteed to make the same decision on the least value received. However, to guarantee a simultaneous decision,
they must still wait until  the existence of a clean round is common knowledge. The condition $\mathtt{current\_waste} \neq t+1 - \mathtt{time}$
is used  to detect the point at which the existence of a clean round is common knowledge.

%ron2: added this section 
\section{Knowledge Based Program - Eventual Byzantine Agreement} \label{sec:kbpeba}

In the Eventual Byzantine Agreement (EBA) problem, the Simultaneous Agreement requirement is replaced by the following property: 
\begin{itemize} 
\item {\bf Agreement(N):}  If $i,j\in N$ and $\decided_i(v)$ and $\decided_j(v')$ then $v=v'$. 
\end{itemize} 
This allows nonfaulty agents to make their decision at different times, but they are still required to agree on the decisions that they make. 

We work with the following knowledge based program $\kbp^0$ for EBA in the sending omissions failures model from \cite{AHM23}, 
which is  similar to a knowledge based program for the crash failures model
in \cite{CGM14,CastanedaMRR17}. 
It is shown in \cite{AHM23} that, subject to some technical side conditions on the information exchange $\exchange$, 
implementations of this knowledge based program are optimal EBA protocols with respect to $\exchange$. 
The side conditions state, in effect, that (1) the only way that an agent can learn that some agent has $\init_i = 0$ 
is through a chain of messages, with some agent deciding 0 at each point of the chain, and (2) if an agent considers it possible that 
some agent is deciding 0, then it considers it possible both that it is nonfaulty itself and 
that some \emph{nonfaulty} agent is deciding 0. Intuitively, these conditions are satisfied
in information exchanges, like the 
%ron3: flooding 
floodSet
protocol, in which agents explicitly transmit information about 0 values that they learn, 
but never store enough information about messages received to be able to make deductions about which agents are faulty. 

\begin{program}
  \DontPrintSemicolon
 \Repeat{$\decided_i$}{
 %\lIf{$\decided_i\neq \bot$}{$\noop$} \lElseIf{
  \lIf{
  $\init_i = 0
    \; \lor \;
            %kaya5: moved \lor inside for consistency
            %\bigvee_{j \in \Agents} K_i (\jdecided_j = 0)$
            K_i (\bigvee_{j \in \Agents} \jdecided_j = 0)$
  }{$\decide_i(0)$}
  \lElseIf{$K_i(\bigwedge_{j \in \Agents} \neg (\deciding_j = 0))$}
  {$\decide_i(1)$}
  \lElse{$\noop$}}
  \caption{$\kbp^0_i$}
\end{program}

Another, more complex, knowledge based program is also considered in  \cite{AHM23}, which does not require these side conditions, 
and also yields an optimal EBA protocol for the sending omissions failure model with respect to the full information exchange. 
This knowledge based program adds some cases to the program $\kbp^0$, expressed using common knowledge, that 
can only be satisfied when the information exchange allows agents to learn which agents are faulty. 
Since we do not consider the full information exchange in the present paper, but only information exchanges that satisfy the 
side conditions for $\kbp^0$, we focus on this simpler protocol. 

%ron2*: Issues: Does the EBA knowledge based program always satisfy the EBA spec? 
%Is it optimal for models other than sending omissions. If not, does the sending omissions optimality result  
%work for at least some other models? 

\section{Information Exchanges for Eventual Byzantine Agreement} \label{sec:infexeba}

For the EBA problem, Alpturer et al.~\cite{AHM23} discuss two information exchanges $\exchange$ that satisfy 
the constraint under which implementations of the knowledge based program $\kbp^0$  are optimal EBA protocols 
with respect to  $\exchange$. 

We describe these exchanges $\exchange^{basic}$ and $\exchange^{min}$ in the present section. 

\newcommand{\rd}{\mathit{jd}}
\newcommand{\numo}{\mathit{num1}}

\subsection{Information Exchange $\exchange^{min}$}

In the information exchange $\exchange^{min}$, agent $i$'s local state is a tuple $\langle\Time_i,\init_i,
    \decided_i,\rd_i\rangle$, where $\Time_i$ is the time, $\init_i$ is the agent's initial value, 
$\decided_i$ records whether the agent has decided, and $\rd_i$, intuitively, 
records either a value that the agent has heard that some agent has just decided, or $\bot$. 
When the agent decides a value $v$, it sends a message comprised just of the value $v$ to all agents. Otherwise, it sends no message.  
When such a value $0$ is received by agent $i$, it sets the value of the variable $\rd_i$ to $0$, 
otherwise, if a message $1$ is received, it sets the value of the variable $\rd_i$ to $1$, 
otherwise the  value of this variable is~$\bot$. 

An implementation of $\kbp^0$ with respect to this information exchange is straightforward. Agent $i$ waits until either 
$\init_i = 0 $ or $\rd_i = 0$, and decides $0$ if this becomes true before time $t+1$. Otherwise, at time $t+1$, the 
agent decides $1$. 

\subsection{Information Exchange $\exchange^{basic}$}

The information exchange $\exchange^{basic}$ is an extension of $\exchange^{min}$
in which agent $i$'s local state is a tuple $\langle\Time_i,\init_i,
    \decided_i,\rd_i,\numo\rangle$.  Here $\Time_i$ is the time, $\init_i$ is the agent's initial value, 
$\decided_i$ records whether the agent has decided,  $\rd_i$ again, records either a value that the agent has just 
heard that some agent has just decided, or $\bot$. The additional variable $\numo$ records a number. 
When the agent decides a value $v$, it sends a message comprised just of the value $v$ to all agents. 
Otherwise, if it has initial value $1$, it sends the message $(\init,1)$, or if it has initial value $0$, it sends no message. 
When such a value $0$ is received by agent $i$, it sets the value of the variable $\rd_i$ to $0$, otherwise the 
value of this variable is $\bot$. In each round, the variable $\numo_i$ is set to
the number of messages of the form $(\init,1)$ 
that the agent has received in the last round. 

It is shown in \cite{AHM23} that this enables an early stopping condition for the EBA problem. An implementation of the knowledge based 
program has agent $i$ deciding 0 when either $\init_i=0$ or $\rd_i = 0$. The agent decides 1 when either 
$\numo_i > n- \Time_i$ or   $\rd_i = 1$.

\section{Model Checking and Synthesis Performance Results} \label{sec:running}

The running times of our model checking and synthesis experiments are reported in this section.
Our experiments were conducted on a machine with
3.7 GHz 6-core Intel Core i5 with 256 KB L2 cache, 9 MB L3 cache and 32 GB memory.
(This machine is multicore, but the present version of MCK runs in a single core.)
The timeout
TO for long running computations
was taken to be
10 minutes. (This was selected to obtain a reasonable runtime for our final full experimental run. In our early testing, a few additional
cases terminated within 5 hours, but many TO entries in the tables below ran for as much as 2 days without termination.)

\subsection{Results for SBA}

Table~\ref{tab:FloodCount} shows the time required to execute the FloodSet and Count FloodSet protocols on $n$ agents, $t$ maximum failures, and the number of values fixed at $v = 2$.

\hspace{-5em}\begin{table}[ht]
    \centering
    \begin{tabular}{SSSSSSS} \toprule
        & & \multicolumn{2}{c}{FloodSet protocol} & \multicolumn{2}{c}{Count FloodSet protocol} \\ \midrule
        {$n$} & {$t$} & {model check} & {synth.} & {model check} & {synth.} \\ \midrule

  {2} & {1} & { 0m0.069} & { 0m0.239} & { 0m0.085} & { 0m0.336} \\
 {2} & {2} & { 0m0.085} & { 0m0.344} & { 0m0.097} & { 0m0.569} \\\midrule

 {3} & {1} & { 0m0.150} & { 0m0.587} & { 0m0.292} & { 0m1.245} \\
 {3} & {2} & { 0m0.228} & { 0m1.011} & { 0m0.407} & { 0m2.401} \\
 {3} & {3} & { 0m0.278} & { 0m1.408} & { 0m0.538} & { 0m4.511} \\ \midrule

 {4} & {1} & { 0m0.672} & { 0m2.706} & { 0m5.274} & { 0m24.263} \\
 {4} & {2} & { 0m2.122} & { 0m8.164} & { 0m23.328} & { 2m0.892} \\
 {4} & {3} & { 0m2.264} & { 1m9.663} & { 0m23.059} & {TO} \\
 {4} & {4} & { 0m3.333} & { 5m40.488} & { 0m27.705} & {TO} \\ \midrule

 {5} & {1} & { 0m6.214} & { 0m25.784} & {TO} & {TO} \\
 {5} & {2} & { 0m34.635} & { 2m42.015} & {TO} & {TO} \\
 {5} & {3} & { 0m41.724} & {TO} & {TO} & {TO}\\
 {5} & {4} & { 1m8.150} & {TO} & {TO} & {TO} \\
 {5} & {5} & { 1m12.180} & {TO} & {TO} & {TO} \\ \midrule

  {6} & {1} & { 1m12.863} & {TO} & {TO} & {TO} \\
  {6} & {2} & { TO} & {TO} & {TO} & {TO} \\

         \bottomrule
    \end{tabular}
     \caption{{\em Running times for number of agents $n$, maximum number of faulty agents $t$}}
    \label{tab:FloodCount}
\end{table}

From these results, we see that adding even a single count variable
has a significant impact on performance, with model checking and synthesis of the counting version of FloodSet  scaling less well, and timing out at a smaller numbers of agents.
Synthesis is more complex than model checking.
A similar poor performance is therefore expected for the
even more complex
Differential Protocol and the Dwork-Moses protocols.
To determine the impact of the number of rounds on the performance, we have considered versions in
which fewer than the required $t+1$ rounds are executed. The results
for model checking
are given in Table~\ref{tab:DiffDM}.

\hspace{-5em}\begin{table}[htp]
    \centering
    \begin{tabular}{SSSSSS} \toprule
    & & & {differential protocol} & {Dwork and Moses} \\ \midrule
       {$n$} & {$t$} & {no. rounds} & {time} & {time} \\ \midrule
 {2} & {1} & {1} & { 0m0.098}  & { 0m0.490} \\
 {2} & {1} & {2}  & { 0m0.104}  & { 0m0.576} \\ \midrule
 {2} & {2} & {1} & { 0m0.110}   & { 0m0.490} \\
 {2} & {2} & {2} & { 0m0.116}  & { 0m0.574}  \\
 {2} & {2} & {3}  & { 0m0.114}  & { 0m0.637} \\ \midrule
 {3} & {1} & {1} & { 4m18.072}  & { TO } \\
 {3} & {1} & {2} & { 5m4.243}  & { TO } \\ \midrule
 {3} & {2} & {1} & { 4m17.531}  & { TO } \\
 {3} & {2} & {2} & { 4m27.026}  & { TO } \\
 {3} & {2} & {3} &  { 4m27.951}  & { TO } \\ \midrule
 {3} & {3} & {1} & { 4m23.998}  & { TO } \\
 {3} & {3} & {2} &  { 4m23.384}  & { TO } \\
 {3} & {3} & {3} & { 4m28.896}  & { TO } \\
 {3} & {3} & {4} & { 4m24.074}  & { TO } \\ \midrule
 {4} & {1} & {1} & {TO} & {TO} \\
         \bottomrule
    \end{tabular}
    \caption{{\em Running times for model checking SBA, Diff and Dwork Moses protocols.}}
    \label{tab:DiffDM}
\end{table}

We see that adding the additional ``previous count'' variable to the Count FloodSet model results in model
checking scaling less well, with models with one fewer agent terminating within the timeout.
However, the number of rounds appears to have a minimal impact on the performance.

\section{Performance Results for EBA}

In
the case of EBA, we report just the results for synthesis.
The optimality result of \cite{AHM23} for the knowledge-based program $\kbp^0$ applies to the  sending omissions model, but this includes the crash
failures model. We have therefore modelled both failures models.

\hspace{-5em}\begin{table}[H]
    \centering
    \begin{tabular}{SSSSSSS} \toprule
    & & \multicolumn{2}{c}{$\exchange^{min}$} & \multicolumn{2}{c}{$\exchange^{basic}$} \\ \midrule
    {$n$} & {$t$} & {crash} & {omissions} & {crash} & {omissions} \\ \midrule
    {2} & {1} & { 0m0.413} & { 0m0.336} & { 0m1.478} & { 0m1.245} \\
    {2} & {2} & { 0m0.623} & { 0m0.408} & { 0m3.651} & { 0m4.976} \\  \midrule
    {3} & {1} & { 0m31.219} & { 0m11.371} & {TO} & {TO} \\
    {3} & {2} & { 2m43.594} & { 0m14.046} & {TO} & {TO} \\
    {3} & {3} & { 2m58.142} & { 1m6.735} & {TO} & {TO} \\ \midrule
    {4} & {1} & {TO} & {TO} & {TO} & {TO} \\
    \bottomrule
    \end{tabular}
    \caption{{\em Running times for EBA synthesis}}
    \label{tab:EBA}
\end{table}

Table \ref{tab:EBA} gives the performance results for synthesis. It appears that, even though the knowledge based program $\kbp^0$ does not involve common knowledge operators, the performance scales less
well than in the SBA case. Whereas for SBA, the FloodSet information exchange scaled to examples with 5 agents, in the case of EBA, the similar information exchange $\exchange^{min}$ scales to just 4 agents before the blowup in computation time. The performance for the information exchange $\exchange^{min}$
is worse. This is expected because, like the Count information exchange considered for SBA, this protocol has an additional variable that counts the number of messages received by an agent.

We have also modelled receiving omissions and general omissions, and the performance results obtained are similar, with successful computations in the
same cases.

\section{Related Work} \label{sec:related}

Knowledge based analyses of consensus protocols have generally focused on full-information protocols, yielding theoretically optimal,
but not necessarily practical implementations \cite{DM90,HalpernMW01,MT88, Moses09}.
Alpturer, Halpern and van der Meyden \cite{AHM23} have previously considered the knowledge based analysis of consensus protocols
and optimality with respect to limited information exchange.
However, the focus of that paper is on Eventual Byzantine Agreement, where agents do not need to decide simultaneously.
The theory underlying knowledge-based programs and the existence of their implementations that we draw upon
is developed in \cite{FHMVbook}.

Automated synthesis of concurrent systems is an established area with its own workshops \cite{synt}.
In general, the focus is on synthesis of protocols from temporal specifications; early works in this
area are  \cite{ALW89,APPG88,AM94, AE89}.
This approach does not guarantee optimality of the resulting protocol in the way that
implementing a knowledge-based program is able to achieve.
There exists some work on synthesis from epistemic specifications
\cite{BensalemPS10,GrafPQ12,KatzPS11}
but in general, the focus is on specifications about a single agent's knowledge, rather than
forms of knowledge involving multiple agents, like the common knowledge condition that
is used in the present paper.

A number of epistemic model checkers exist \cite{GammieM04,LomuscioQR09,verics}, but
MCK remains the only such system to address the automated synthesis of implementations of knowledge-based programs.  Of the range of problems that have been studied using epistemic model checking,
closest to the SBA problem we consider here is a work by Al-Bateineh and Reynolds \cite{Al-BatainehR19}
that considers the Byzantine Atomic Commitment problem. Synthesis is not attempted in this work.

 \section{Conclusion} \label{sec:concl}

In this paper, we have investigated the application of epistemic model checking and synthesis
simultaneous and eventual versions of Byzantine Agreement.
We have shown that this technique,
at least on some small examples, is able to automatically identify situations where the information being exchanged
in a protocol provides an opportunity to make a decision earlier than in the protocol as originally designed,
leading to  implementations that are optimal relative to that information exchange.

We find in our experiments that synthesis scales less well than model checking.
Methodologically, this suggests that epistemic synthesis can be used on small scale instances, and the results used to developed a hypothesis concerning the
predicates for optimal termination.
Model checking can then be used on slightly larger instances, to test this hypothesis. Having developed a hypothesis that survives verification by model checking, a manual proof of correctness and optimality will still need to be developed
to obtain a result for all numbers of agents and failures.

One disappointing aspect of our results is that both model checking and synthesis
time out at a low number of agents, with a dramatic blowup of performance at the threshold.
To some extent, a blowup is expected, since the
 information exchanges we have considered inherently involve $n(n-1)$ messages per round.
The model checking algorithms that we use for this exercise use Binary Decision Diagrams,
which tend to scale to models involving just 100-200 variables in the boolean representation.

Temporal model checking usually scales better using SAT-based bounded model checking techniques,
but because we need to check a complex common knowledge fixpoint, and negative occurrences
of knowledge operators, this technique cannot be applied to our problem.
However, we remark that we are able to model check the purely temporal specification of SBA
using MCK's SAT-based model checking, with significantly better scaling. For example,
model checking the SBA specification on the $n=5$, $t=4$ model of the Dwork-Moses protocol takes just  2 minutes 5 seconds.

Nevertheless,  our results demonstrate in principle feasibility, and motivate future research aimed at
improving upon the benchmarks for epistemic model checking and synthesis set in our experiments.

Various directions for future work suggest themselves from this. One is to consider
consensus protocols with linear messaging - conceivably, the BDD blowup experienced in the
present study will be ameliorated in such protocols. Another is to develop
alternate epistemic model checking and synthesis algorithms that scale better on these examples.
Of particular interest would be parametric epistemic model checking and synthesis techniques that
address the issue of inherent quadratic scaling of the model checking inputs. As we have written
generators for arbitrary size models, one of the outcomes of the present work is a set of challenge problems for research on
epistemic model checkers.

\bibliography{bib}


\begin{thebibliography}{39}

%%% the Web address.

\ifx \showCODEN    \undefined \def \showCODEN     #1{\unskip}     \fi
\ifx \showDOI      \undefined \def \showDOI       #1{#1}\fi
\ifx \showISBNx    \undefined \def \showISBNx     #1{\unskip}     \fi
\ifx \showISBNxiii \undefined \def \showISBNxiii  #1{\unskip}     \fi
\ifx \showISSN     \undefined \def \showISSN      #1{\unskip}     \fi
\ifx \showLCCN     \undefined \def \showLCCN      #1{\unskip}     \fi
\ifx \shownote     \undefined \def \shownote      #1{#1}          \fi
\ifx \showarticletitle \undefined \def \showarticletitle #1{#1}   \fi
\ifx \showURL      \undefined \def \showURL       {\relax}        \fi
% invisible to TeX
\providecommand\bibfield[2]{#2}
\providecommand\bibinfo[2]{#2}
\providecommand\natexlab[1]{#1}
\providecommand\showeprint[2][]{arXiv:#2}

\bibitem[syn(2024)]%
        {synt}
 \bibinfo{year}{2012-2024}\natexlab{}.
\newblock \bibinfo{booktitle}{\emph{SYNT workshop series}}.
\newblock
\newblock
\shownote{\url{https://cgi.csc.liv.ac.uk/~sven/synt/}}.


\bibitem[Abadi et~al\mbox{.}(1989)]%
        {ALW89}
\bibfield{author}{\bibinfo{person}{M. Abadi}, \bibinfo{person}{L. Lamport},
  {and} \bibinfo{person}{P. Wolper}.} \bibinfo{year}{1989}\natexlab{}.
\newblock \showarticletitle{Realizable and Unrealizable Concurrent Program
  Specifications}. In \bibinfo{booktitle}{\emph{Proc. 16th Int. Colloquium on
  Automata, Languages and Programming}}, Vol.~\bibinfo{volume}{372}.
  \bibinfo{publisher}{Lecture Notes in Computer Science, Springer-Verlag},
  \bibinfo{pages}{1--17}.
\newblock


\bibitem[Afrati et~al\mbox{.}(1986)]%
        {APPG88}
\bibfield{author}{\bibinfo{person}{F. Afrati}, \bibinfo{person}{C.
  Papadimitriou}, {and} \bibinfo{person}{G. Papageorgiou}.}
  \bibinfo{year}{1986}\natexlab{}.
\newblock \showarticletitle{The synthesis of communication protocols}. In
  \bibinfo{booktitle}{\emph{PODC '86: Proc. 5th ACM symposium on Principles of
  Distributed Computing}}. \bibinfo{pages}{263--271}.
\newblock


\bibitem[Al{-}Bataineh and Reynolds(2019)]%
        {Al-BatainehR19}
\bibfield{author}{\bibinfo{person}{O.~I. Al{-}Bataineh} {and}
  \bibinfo{person}{M. Reynolds}.} \bibinfo{year}{2019}\natexlab{}.
\newblock \showarticletitle{Epistemic model checking of distributed commit
  protocols with byzantine faults}. In \bibinfo{booktitle}{\emph{Proceedings of
  the 7th International Workshop on Formal Methods in Software Engineering,
  FormaliSE@ICSE 2019, Montreal, QC, Canada, May 27, 2019}},
  \bibfield{editor}{\bibinfo{person}{Stefania Gnesi}, \bibinfo{person}{Nico
  Plat}, \bibinfo{person}{Nancy~A. Day}, {and} \bibinfo{person}{Matteo Rossi}}
  (Eds.). \bibinfo{publisher}{{IEEE} / {ACM}}, \bibinfo{pages}{51--60}.
\newblock
\urldef\tempurl%
\url{https://doi.org/10.1109/FORMALISE.2019.00014}
\showDOI{\tempurl}


\bibitem[Al{-}Bataineh and van~der Meyden(2010)]%
        {Al-BatainehM10}
\bibfield{author}{\bibinfo{person}{O.~I. Al{-}Bataineh} {and}
  \bibinfo{person}{R. van~der Meyden}.} \bibinfo{year}{2010}\natexlab{}.
\newblock \showarticletitle{Epistemic Model Checking for Knowledge-Based
  Program Implementation: An Application to Anonymous Broadcast}. In
  \bibinfo{booktitle}{\emph{Proc. SecureComm}}. \bibinfo{pages}{429--447}.
\newblock


\bibitem[Alpturer et~al\mbox{.}(2023)]%
        {AHM23}
\bibfield{author}{\bibinfo{person}{K. Alpturer}, \bibinfo{person}{J.~Y.
  Halpern}, {and} \bibinfo{person}{R. van~der Meyden}.}
  \bibinfo{year}{2023}\natexlab{}.
\newblock \showarticletitle{Optimal Eventual Byzantine Agreement Protocols with
  Omission Failures}. In \bibinfo{booktitle}{\emph{Proc. {ACM} Symp. on
  Principles of Distributed Computing, {PODC}}}. \bibinfo{publisher}{{ACM}},
  \bibinfo{pages}{244--252}.
\newblock
\urldef\tempurl%
\url{https://doi.org/10.1145/3583668.3594573}
\showDOI{\tempurl}


\bibitem[Alpturer et~al\mbox{.}(2025)]%
        {AMRW24}
\bibfield{author}{\bibinfo{person}{K. Alpturer}, \bibinfo{person}{R. van~der
  Meyden}, \bibinfo{person}{S. Ruj}, {and} \bibinfo{person}{G. Wong}.}
  \bibinfo{year}{2025}\natexlab{}.
\newblock \bibinfo{title}{Optimality of Simultaneous Byzantine Agreement with
  Limited Information Exchange}.  (\bibinfo{year}{2025}).
\newblock
\newblock
\shownote{to appear, Conf. on Theoretical Aspects of Rationality and Knowledge,
  TARK, July 2025}.


\bibitem[Anuchitanukul and Manna(1994)]%
        {AM94}
\bibfield{author}{\bibinfo{person}{A. Anuchitanukul} {and} \bibinfo{person}{Z.
  Manna}.} \bibinfo{year}{1994}\natexlab{}.
\newblock \showarticletitle{Realizability and Synthesis of Reactive Modules}.
  In \bibinfo{booktitle}{\emph{Computer-Aided Verification, Proc. 6th Int'l
  Conference}}. \bibinfo{publisher}{Springer-Verlag, Lecture Notes in Computer
  Science 818}, \bibinfo{address}{Stanford, California},
  \bibinfo{pages}{156--169}.
\newblock


\bibitem[Attie and Emerson(1989)]%
        {AE89}
\bibfield{author}{\bibinfo{person}{P.~C. Attie} {and} \bibinfo{person}{E.A.
  Emerson}.} \bibinfo{year}{1989}\natexlab{}.
\newblock \showarticletitle{Synthesis of Concurrent Systems with Many Similar
  Sequential Processes}. In \bibinfo{booktitle}{\emph{Proc. 16th {ACM}
  Symposium on Principles of Programming Languages}}.
  \bibinfo{address}{Austin}, \bibinfo{pages}{191--201}.
\newblock


\bibitem[Baukus and van~der Meyden(2004)]%
        {BaukusM04}
\bibfield{author}{\bibinfo{person}{K. Baukus} {and} \bibinfo{person}{R. van~der
  Meyden}.} \bibinfo{year}{2004}\natexlab{}.
\newblock \showarticletitle{A Knowledge Based Analysis of Cache Coherence}. In
  \bibinfo{booktitle}{\emph{Proc. ICFEM}}. \bibinfo{pages}{99--114}.
\newblock


\bibitem[Bensalem et~al\mbox{.}(2010)]%
        {BensalemPS10}
\bibfield{author}{\bibinfo{person}{S. Bensalem}, \bibinfo{person}{D. Peled},
  {and} \bibinfo{person}{J. Sifakis}.} \bibinfo{year}{2010}\natexlab{}.
\newblock \showarticletitle{Knowledge Based Scheduling of Distributed Systems}.
  In \bibinfo{booktitle}{\emph{Time for Verification, Essays in Memory of Amir
  Pnueli}} \emph{(\bibinfo{series}{Lecture Notes in Computer Science},
  Vol.~\bibinfo{volume}{6200})}. \bibinfo{publisher}{Springer},
  \bibinfo{pages}{26--41}.
\newblock


\bibitem[Burch et~al\mbox{.}(1992)]%
        {BCDHM90}
\bibfield{author}{\bibinfo{person}{J.~R. Burch}, \bibinfo{person}{E.~M.
  Clarke}, \bibinfo{person}{D.~L. Dill}, \bibinfo{person}{J. Hwang}, {and}
  \bibinfo{person}{K.~L. McMillan}.} \bibinfo{year}{1992}\natexlab{}.
\newblock \showarticletitle{Symbolic model checking: $10^{20}$ states and
  beyond}.
\newblock \bibinfo{journal}{\emph{Information and Computation}}
  \bibinfo{volume}{98}, \bibinfo{number}{2} (\bibinfo{year}{1992}),
  \bibinfo{pages}{142--171}.
\newblock


\bibitem[Casta{\~{n}}eda et~al\mbox{.}(2014)]%
        {CGM14}
\bibfield{author}{\bibinfo{person}{A. Casta{\~{n}}eda}, \bibinfo{person}{Y.~A.
  Gonczorowski}, {and} \bibinfo{person}{Y. Moses}.}
  \bibinfo{year}{2014}\natexlab{}.
\newblock \showarticletitle{Unbeatable consensus}. In
  \bibinfo{booktitle}{\emph{Proc.~28th International Conference on Distributed
  Computing (DISC '14)}}. \bibinfo{pages}{91--106}.
\newblock


\bibitem[Casta{\~{n}}eda et~al\mbox{.}(2017)]%
        {CastanedaMRR17}
\bibfield{author}{\bibinfo{person}{A. Casta{\~{n}}eda}, \bibinfo{person}{Y.
  Moses}, \bibinfo{person}{M. Raynal}, {and} \bibinfo{person}{M. Roy}.}
  \bibinfo{year}{2017}\natexlab{}.
\newblock \showarticletitle{Early Decision and Stopping in Synchronous
  Consensus: {A} Predicate-Based Guided Tour}. In
  \bibinfo{booktitle}{\emph{Networked Systems - 5th International Conference,
  {NETYS} 2017, Proc.}} \bibinfo{pages}{206--221}.
\newblock
\urldef\tempurl%
\url{https://doi.org/10.1007/978-3-319-59647-1_16}
\showDOI{\tempurl}


\bibitem[Clarke et~al\mbox{.}(1999)]%
        {clarkebook}
\bibfield{author}{\bibinfo{person}{E. Clarke}, \bibinfo{person}{O. Grumberg},
  {and} \bibinfo{person}{D. Peled}.} \bibinfo{year}{1999}\natexlab{}.
\newblock \bibinfo{booktitle}{\emph{{Model Checking}}}.
\newblock \bibinfo{publisher}{The MIT Press}.
\newblock
\showISBNx{0262032708}


\bibitem[Dembinski et~al\mbox{.}(2003)]%
        {verics}
\bibfield{author}{\bibinfo{person}{P. Dembinski}, \bibinfo{person}{A.
  Janowska}, \bibinfo{person}{P. Janowski}, \bibinfo{person}{W. Penczek},
  \bibinfo{person}{A. P{\'o}lrola}, \bibinfo{person}{M. Szreter},
  \bibinfo{person}{B. Wozna}, {and} \bibinfo{person}{A. Zbrzezny}.}
  \bibinfo{year}{2003}\natexlab{}.
\newblock \showarticletitle{Verics: A Tool for Verifying Timed Automata and
  {Estelle} Specifications}. In \bibinfo{booktitle}{\emph{Proc. Conf. Tools and
  Algorithms for the Construction and Analysis of Systems}}.
  \bibinfo{pages}{278--283}.
\newblock


\bibitem[Dolev et~al\mbox{.}(1990)]%
        {DolevRS90}
\bibfield{author}{\bibinfo{person}{D. Dolev}, \bibinfo{person}{R. Reischuk},
  {and} \bibinfo{person}{H.~R. Strong}.} \bibinfo{year}{1990}\natexlab{}.
\newblock \showarticletitle{Early Stopping in Byzantine Agreement}.
\newblock \bibinfo{journal}{\emph{J. {ACM}}} \bibinfo{volume}{37},
  \bibinfo{number}{4} (\bibinfo{year}{1990}), \bibinfo{pages}{720--741}.
\newblock


\bibitem[Dwork and Moses(1990)]%
        {DM90}
\bibfield{author}{\bibinfo{person}{C. Dwork} {and} \bibinfo{person}{Y. Moses}.}
  \bibinfo{year}{1990}\natexlab{}.
\newblock \showarticletitle{Knowledge and common knowledge in a {B}yzantine
  environment: crash failures}.
\newblock \bibinfo{journal}{\emph{Information and Computation}}
  \bibinfo{volume}{88}, \bibinfo{number}{2} (\bibinfo{year}{1990}),
  \bibinfo{pages}{156--186}.
\newblock


\bibitem[Fagin et~al\mbox{.}(2003)]%
        {FHMVbook}
\bibfield{author}{\bibinfo{person}{R. Fagin}, \bibinfo{person}{J.Y. Halpern},
  \bibinfo{person}{Y. Moses}, {and} \bibinfo{person}{M.Y. Vardi}.}
  \bibinfo{year}{2003}\natexlab{}.
\newblock \bibinfo{booktitle}{\emph{{Reasoning About Knowledge}}}.
\newblock \bibinfo{publisher}{The MIT Press}.
\newblock


\bibitem[Gammie and {van der Meyden}(2004)]%
        {GammieM04}
\bibfield{author}{\bibinfo{person}{P. Gammie} {and} \bibinfo{person}{R. {van
  der Meyden}}.} \bibinfo{year}{2004}\natexlab{}.
\newblock \showarticletitle{{MCK}: Model Checking the Logic of Knowledge.}. In
  \bibinfo{booktitle}{\emph{Proc. Conf. on Computer Aided Verification (CAV)}}
  \emph{(\bibinfo{series}{LNCS}, Vol.~\bibinfo{volume}{3114})},
  \bibfield{editor}{\bibinfo{person}{Rajeev Alur} {and} \bibinfo{person}{Doron
  Peled}} (Eds.). \bibinfo{publisher}{Springer}, \bibinfo{pages}{479--483}.
\newblock


\bibitem[Graf et~al\mbox{.}(2012)]%
        {GrafPQ12}
\bibfield{author}{\bibinfo{person}{S. Graf}, \bibinfo{person}{D. Peled}, {and}
  \bibinfo{person}{S. Quinton}.} \bibinfo{year}{2012}\natexlab{}.
\newblock \showarticletitle{Achieving distributed control through model
  checking}.
\newblock \bibinfo{journal}{\emph{Formal Methods in System Design}}
  \bibinfo{volume}{40}, \bibinfo{number}{2} (\bibinfo{year}{2012}),
  \bibinfo{pages}{263--281}.
\newblock


\bibitem[Halpern and Moses(1990)]%
        {HalpernM90}
\bibfield{author}{\bibinfo{person}{J.~Y. Halpern} {and} \bibinfo{person}{Y.
  Moses}.} \bibinfo{year}{1990}\natexlab{}.
\newblock \showarticletitle{Knowledge and Common Knowledge in a Distributed
  Environment}.
\newblock \bibinfo{journal}{\emph{J. {ACM}}} \bibinfo{volume}{37},
  \bibinfo{number}{3} (\bibinfo{year}{1990}), \bibinfo{pages}{549--587}.
\newblock


\bibitem[Halpern et~al\mbox{.}(2001)]%
        {HalpernMW01}
\bibfield{author}{\bibinfo{person}{J.~Y. Halpern}, \bibinfo{person}{Y. Moses},
  {and} \bibinfo{person}{O. Waarts}.} \bibinfo{year}{2001}\natexlab{}.
\newblock \showarticletitle{A Characterization of Eventual Byzantine
  Agreement}.
\newblock \bibinfo{journal}{\emph{{SIAM} J. Comput.}} \bibinfo{volume}{31},
  \bibinfo{number}{3} (\bibinfo{year}{2001}), \bibinfo{pages}{838--865}.
\newblock


\bibitem[Huang et~al\mbox{.}(2011)]%
        {HuangMM11}
\bibfield{author}{\bibinfo{person}{X. Huang}, \bibinfo{person}{P. Maupin},
  {and} \bibinfo{person}{R. van~der Meyden}.} \bibinfo{year}{2011}\natexlab{}.
\newblock \showarticletitle{Model Checking Knowledge in Pursuit Evasion Games}.
  In \bibinfo{booktitle}{\emph{Proc. IJCAI}}. \bibinfo{pages}{240--245}.
\newblock


\bibitem[Huang and van~der Meyden(2013a)]%
        {HMtark13}
\bibfield{author}{\bibinfo{person}{X. Huang} {and} \bibinfo{person}{R. van~der
  Meyden}.} \bibinfo{year}{2013}\natexlab{a}.
\newblock \showarticletitle{Symbolic Synthesis of Knowledge-based Program
  Implementations with Synchronous Semantics}. In
  \bibinfo{booktitle}{\emph{Proc. TARK}}. \bibinfo{pages}{121--130}.
\newblock


\bibitem[Huang and van~der Meyden(2013b)]%
        {HuangM13}
\bibfield{author}{\bibinfo{person}{X. Huang} {and} \bibinfo{person}{R. van~der
  Meyden}.} \bibinfo{year}{2013}\natexlab{b}.
\newblock \showarticletitle{Symbolic Synthesis of Knowledge-based Program
  Implementations with Synchronous Semantics}. In
  \bibinfo{booktitle}{\emph{Proc. TARK}}.
\newblock


\bibitem[Huang and van~der Meyden(2014a)]%
        {HM14tacas}
\bibfield{author}{\bibinfo{person}{X. Huang} {and} \bibinfo{person}{R. van~der
  Meyden}.} \bibinfo{year}{2014}\natexlab{a}.
\newblock \showarticletitle{Symbolic Synthesis for Epistemic Specifications
  with Observational Semantics}. In \bibinfo{booktitle}{\emph{Proc. Tools and
  Algorithms for the Construction and Analysis of Systems, {TACAS}}}.
  \bibinfo{pages}{455--469}.
\newblock
\urldef\tempurl%
\url{https://doi.org/10.1007/978-3-642-54862-8_39}
\showDOI{\tempurl}


\bibitem[Huang and van~der Meyden(2014b)]%
        {HuangM14}
\bibfield{author}{\bibinfo{person}{X. Huang} {and} \bibinfo{person}{R. van~der
  Meyden}.} \bibinfo{year}{2014}\natexlab{b}.
\newblock \showarticletitle{Symbolic Synthesis for Epistemic Specifications
  with Observational Semantics}. In \bibinfo{booktitle}{\emph{Proc. TACAS}}.
  \bibinfo{pages}{455--469}.
\newblock


\bibitem[Huang and van~der Meyden(2015)]%
        {HuangM16}
\bibfield{author}{\bibinfo{person}{X. Huang} {and} \bibinfo{person}{R. van~der
  Meyden}.} \bibinfo{year}{2015}\natexlab{}.
\newblock \showarticletitle{The complexity of approximations for epistemic
  synthesis (extended abstract)}. In \bibinfo{booktitle}{\emph{Proc. Workshop
  on Synthesis, {SYNT}}}. \bibinfo{pages}{120--137}.
\newblock


\bibitem[Katz et~al\mbox{.}(2011)]%
        {KatzPS11}
\bibfield{author}{\bibinfo{person}{G. Katz}, \bibinfo{person}{D. Peled}, {and}
  \bibinfo{person}{S. Schewe}.} \bibinfo{year}{2011}\natexlab{}.
\newblock \showarticletitle{Synthesis of Distributed Control through Knowledge
  Accumulation}. In \bibinfo{booktitle}{\emph{Proc. Int. Conf on Computer Aided
  Verification}}. \bibinfo{pages}{510--525}.
\newblock


\bibitem[Lomuscio et~al\mbox{.}(2009)]%
        {LomuscioQR09}
\bibfield{author}{\bibinfo{person}{A. Lomuscio}, \bibinfo{person}{H. Qu}, {and}
  \bibinfo{person}{F. Raimondi}.} \bibinfo{year}{2009}\natexlab{}.
\newblock \showarticletitle{{MCMAS}: A Model Checker for the Verification of
  Multi-Agent Systems.}. In \bibinfo{booktitle}{\emph{Proc. Conf. on Computer
  Aided Verification (CAV)}} \emph{(\bibinfo{series}{Lecture Notes in Computer
  Science}, Vol.~\bibinfo{volume}{5643})},
  \bibfield{editor}{\bibinfo{person}{Ahmed Bouajjani} {and}
  \bibinfo{person}{Oded Maler}} (Eds.). \bibinfo{publisher}{Springer},
  \bibinfo{pages}{682--688}.
\newblock


\bibitem[Lynch(1996)]%
        {lynch}
\bibfield{author}{\bibinfo{person}{N. Lynch}.} \bibinfo{year}{1996}\natexlab{}.
\newblock \bibinfo{booktitle}{\emph{Distributed Algorithms}}.
\newblock \bibinfo{publisher}{MIT Press}.
\newblock


\bibitem[Meyden(2024)]%
        {LIOPT}
\bibfield{author}{\bibinfo{person}{{R. van der} Meyden}.}
  \bibinfo{year}{2024}\natexlab{}.
\newblock \bibinfo{title}{Optimal Simultaneous Byzantine Agreement, Common
  Knowledge and Limited Information Exchange}.
\newblock
\newblock
\urldef\tempurl%
\url{https://cgi.cse.unsw.edu.au/~meyden/research/common-knowledge.pdf}
\showURL{%
\tempurl}


\bibitem[Meyden and Su(2004)]%
        {MeydenS04}
\bibfield{author}{\bibinfo{person}{{R. van der} Meyden} {and}
  \bibinfo{person}{K. Su}.} \bibinfo{year}{2004}\natexlab{}.
\newblock \showarticletitle{Symbolic Model Checking the Knowledge of the Dining
  Cryptographers}. In \bibinfo{booktitle}{\emph{Proc. CSFW}}.
  \bibinfo{pages}{280--291}.
\newblock


\bibitem[Moses(2009)]%
        {Moses09}
\bibfield{author}{\bibinfo{person}{Y. Moses}.} \bibinfo{year}{2009}\natexlab{}.
\newblock \showarticletitle{Optimum Simultaneous Consensus for General
  Omissions Is Equivalent to an {NP} Oracle}. In
  \bibinfo{booktitle}{\emph{Proc. Distributed Computing, 23rd International
  Symposium, {DISC} 2009}} \emph{(\bibinfo{series}{Lecture Notes in Computer
  Science}, Vol.~\bibinfo{volume}{5805})},
  \bibfield{editor}{\bibinfo{person}{I.~Keidar}} (Ed.).
  \bibinfo{publisher}{Springer}, \bibinfo{pages}{436--448}.
\newblock
\urldef\tempurl%
\url{https://doi.org/10.1007/978-3-642-04355-0_45}
\showDOI{\tempurl}


\bibitem[Moses and Tuttle(1988)]%
        {MT88}
\bibfield{author}{\bibinfo{person}{Y. Moses} {and} \bibinfo{person}{M.~R.
  Tuttle}.} \bibinfo{year}{1988}\natexlab{}.
\newblock \showarticletitle{Programming simultaneous actions using common
  knowledge}.
\newblock \bibinfo{journal}{\emph{Algorithmica}}  \bibinfo{volume}{3}
  (\bibinfo{year}{1988}), \bibinfo{pages}{121--169}.
\newblock
\urldef\tempurl%
\url{https://doi.org/10.1007/BF01762112}
\showDOI{\tempurl}


\bibitem[Pease et~al\mbox{.}(1980)]%
        {PSL}
\bibfield{author}{\bibinfo{person}{M. Pease}, \bibinfo{person}{R. Shostak},
  {and} \bibinfo{person}{L. Lamport}.} \bibinfo{year}{1980}\natexlab{}.
\newblock \showarticletitle{Reaching agreement in the presence of faults}.
\newblock \bibinfo{journal}{\emph{J. ACM}} \bibinfo{volume}{27},
  \bibinfo{number}{2} (\bibinfo{year}{1980}), \bibinfo{pages}{228--234}.
\newblock


\bibitem[Team({[n.\,d.]})]%
        {mckManual}
\bibfield{author}{\bibinfo{person}{MCK Team}.}
  \bibinfo{year}{[n.\,d.]}\natexlab{}.
\newblock \bibinfo{booktitle}{\emph{{MCK User Manual}}}.
\newblock
\newblock
\shownote{Available from \url{http://www.cse.unsw.edu.au/~mck}}.


\bibitem[Vardi(1988)]%
        {Vardi88}
\bibfield{author}{\bibinfo{person}{M.~Y. Vardi}.}
  \bibinfo{year}{1988}\natexlab{}.
\newblock \showarticletitle{A temporal fixpoint calculus}. In
  \bibinfo{booktitle}{\emph{Proc. ACM Symp. on Principles of Programming
  Languages, POPL}}. \bibinfo{pages}{250--259}.
\newblock


\end{thebibliography}

\appendix

\begin{arxiv}
\section{An example of an MCK Synthesis Script} \label{app:floodset}

The following is an example of an MCK synthesis script for the FloodSet protocol,
for $n=3$ agents, $t=1$ possible failures, and $k=2$ values.

The script begins with a statement selecting  the semantics for knowledge to be used in the synthesis process. We used the clock semantics $\mathtt{clk}$ in order to
ensure that there is a unique implementation of the knowledge-based program. In this semantics, an agent's local state consists just of the variables it observes,
as well as the clock value (number of rounds). This is followed by a declaration of some enumerated types,
and a declaration of variables in the environment of the agents. For efficiency of the model,
we have chosen to model the information exchange protocol, including the
way that local states are updated, as part of the environment, rather than in the agents' protocol section.

\begin{minted}[tabsize=2,breaklines]{haskell}
KBP_semantics = clk

type Crash_Status = {ALIVE, CRASHING, CRASHED}
type Time = {0..3}
type Values = {0..1}

vote : Values[Agent]
time : Time
w: Bool[Agent][Values]
--old_w stores value of w at the start of a transition
old_w: Bool[Agent][Values]
status : Crash_Status[Agent]
max_crashed : Time
crashed : Time
\end{minted}

The variable $\mathtt{vote}$ is an array of values indexed by agents; $\mathtt{vote}[i]$ represents the initial preference of agent $i$.
Next, the $\mathtt{init\_cond}$ statement gives a formula that constrains the  possible initial values of these variables.

\begin{minted}[tabsize=2,breaklines]{haskell}
init_cond =
time == 0 /\ max_crashed == 1 /\ crashed == 0 /\
Forall i:Agent (Forall v:Values ( (w[i][v] <=> vote[i] == v) /\ neg old_w[i][v])) /\
Forall i:Agent ( status[i] == ALIVE )
\end{minted}

This is followed by a declaration of the three agents in the system, indicating (in quotes) which protocol each runs,
and the binding of the parameters of the protocol to variables in the environment.

\begin{minted}[tabsize=2,breaklines]{haskell}
agent D0 "decider" (status[D0],time,w[D0])
agent D1 "decider" (status[D1],time,w[D1])
agent D2 "decider" (status[D2],time,w[D2])
\end{minted}

Next, we have a $\mathtt{transitions}$ statement giving code describing how the environment variables are updated in each round.
The statement form \verb+[[ var | formula(var,var' )]]+ is a refinement calculus/TLA action style statement,
meaning that \verb+var+ are the only variables that may change when this statement runs, and its effect is such that
the new values \verb+var'+ after its execution are related to the values \verb+var+ at the start of the execution by the given formula.
This statement allows nondeterminism to be represented. Similarly, the \verb+if+ construct is a Dijkstra style nondeterministic guarded statement.
If more than one guard holds, any corresponding branch may be taken. An example of this is when $\mathtt{status[j]=CRASHING}$, when either agent $\mathtt{j}$'s message
maybe either delivered (updating $\mathtt{w[i]}$), or not (by choosing the $\mathtt{skip}$ branch).

\begin{minted}[tabsize=2,breaklines]{haskell}
transitions
begin
if time < 2 -> begin
-- make a copy, the values to be transmitted
[[ old_w | Forall i:Agent (Forall v:Values ( old_w[i][v]' <=> w[i][v] )) ]];
-- select the agents that crash, keeping the total number crashed at most max_crashed
for i in Agent do
  begin
    [[ status[i] |status[i]' in {ALIVE, CRASHING, CRASHED}  /\
                  (status[i]' == CRASHING => crashed < max_crashed ) /\
                  (status[i] == CRASHED <=> status[i]' == CRASHED) ]];
    if status[i] == CRASHING then crashed := crashed + 1 else skip
  end ;
for i in Agent do
  for j in Agent do
     if status[j] in {ALIVE, CRASHING} ->
            for v in Values do w[i][v] := w[i][v] \/ old_w[j][v]
       [] status[j] in {CRASHING, CRASHED} -> skip
    fi;
for i in Agent do if status[i] == CRASHING -> status[i] := CRASHED fi
end fi ;
-- wipe out the old values, to remove correlations in the BDD
[[ old_w | Forall i:Agent (Forall v:Values (neg old_w[i][v]'))  ]];
time := time + 1
end
\end{minted}

The following are the formulas that we verify after synthesis has been performed. These include formulas for the
temporal specification for SBA, as well as epistemic formulas that capture the situations in which the agent has common knowledge.
The temporal operators used here are from the branching time logic CTL. The operator $\mathtt{A}$ represents ``on all branches'', the
operator $\mathtt{G}$ means ``at all future times'', $\mathtt{X}$ means ``at the next moment of time'' (after the next round) and exponentiation
is used to repeat an operator a give number of times. The keyword ``$\mathtt{spec\_obs}$'' indicates that the observational semantics should be used
to interpret the knowledge operators, but since the time is one of the observable variables, this is equivalent to the clock semantics.

\begin{minted}[tabsize=2,breaklines]{haskell}
spec_obs =
   "Agreement: no conflicting decisions by non-failed agents"
     AG ( Forall i:Agent:"decider" (Forall j:Agent:"decider" (
        (status[i] /= CRASHED /\ i.decided /\
         status[j] /= CRASHED /\ j.decided) =>
               (i.decision == j.decision))))

spec_obs = "Uniform Agreement: all agents that decide agree"
     AG ( Forall i: Agent:"decider" (Forall j:Agent:"decider" (
             (i.decided /\ j.decided) => (i.decision == j.decision) )))

spec_obs =
   "Strong Validity: any decision value is the initial vote of some agent"
   AX^3 ( Forall i:Agent:"decider" (Forall v:Values (
       (i.decision == v /\ i.decided) => Exists j:Agent (vote[j] == v))))

spec_obs =
    "Termination: all nonfaulty agents eventually decide"
   AX^3 ( Forall i:Agent:"decider" (status[i] == ALIVE => i.decided))

spec_obs =
    "agent D0's knowledge test for deciding D0 never holds at time 1"
    AX^1 neg Knows D0 (status[D0] == ALIVE =>
         (gfp _X (Forall i:Agent (status[i] == ALIVE =>
                  (Knows i  (status[i] == ALIVE =>
                          ( (Exists j:Agent (vote[j] == 0)) /\ _X )))))))

spec_obs =
   "at time 2, agent D0's knowledge test for deciding 0 is equivalent to the test used by agent D0"
    AX^2 (D0.values_received[0] <=> Knows D0 (status[D0] == ALIVE =>
          (gfp _X (Forall i:Agent (status[i] == ALIVE =>
                  (Knows i  (status[i] == ALIVE =>
                       ( (Exists j:Agent (vote[j] == 0)) /\ _X ))))))))
\end{minted}

Finally, we have the knowledge based program run by the agents. The variables in the parameters of the protocol are aliased to
environment variables in the agent declarations above. Variables that are declared $\mathtt{observable}$ make up the
agent's local state for purposes of determining what an agent knows. Variables declared $\mathtt{template}$ are boolean variables that
correspond to ``holes'' in the protocol that are to be filled by the synthesizer. These variables occur in conditions in the conditional statements
in the protocol code. The $\mathtt{define}$ statements are macros defining some abbreviations. These are used in the $\mathtt{require}$ statements,
which state properties that must be satisfied by the concrete predicates over the observable variables that are constructed by the synthesizer. In the case of
this knowledge based program, these properties state that each template variable is equivalent to the agent's knowledge that there is common knowledge amongst the
 nonfailed ($status = \mathtt{ALIVE}$) agents. The operator $\mathtt{X}$ in these formulas is the linear temporal logic ``at the next time'' operator.

\begin{minted}[tabsize=2,breaklines]{haskell}
protocol "decider"(status: Crash_Status,
                        time: observable Time,
                        values_received : observable Bool[Values])

decision : Values
decided : Bool

c_1_0 : template
c_1_1 : template
c_2_0 : template
c_2_1 : template



init_cond = neg decided

define someone_voted0 = Env.vote[D0] == 0 \/ Env.vote[D1] == 0 \/ Env.vote[D2] == 0

define someone_voted1 = Env.vote[D0] == 1 \/ Env.vote[D1] == 1 \/ Env.vote[D2] == 1

define decide_condition0 = Knows Self (status == ALIVE => (gfp _X (
       Forall i:Agent:"decider" (i.status == ALIVE => Knows i (i.status == ALIVE => someone_voted0 /\ _X)))))

define decide_condition1 = Knows Self (status == ALIVE => (gfp _X (
       Forall i:Agent:"decider" (i.status == ALIVE => Knows i (i.status == ALIVE => someone_voted1 /\ _X)))))

require = X^1(c_1_0 <=> decide_condition0)
require = X^1(c_1_1 <=> decide_condition1)
require = X^2(c_2_0 <=> decide_condition0)
require = X^2(c_2_1 <=> decide_condition1)

begin
skip;
if (status /= CRASHED /\ neg decided) ->
if c_1_0 then <| decision := 0; decided := True |>  else
if c_1_1 then <| decision := 1; decided := True |>  else
skip
fi
;
if (status /= CRASHED /\ neg decided) ->
if c_2_0 then <| decision := 0; decided := True |>  else
if c_2_1 then <| decision := 1; decided := True |>  else
skip
fi
end

\end{minted}

When MCK runs on this input script, it synthesizes predicates over the observable variables for each of the
template variables, and replaces the template variable statements by the following definition statements.
These statements contain a disjunct for each agent that is running the knowledge based program.
In general, the synthesis result can be different for each agent, but because the present situation is symmetric,
the same predicate is synthesized for each agent.

\begin{minted}[tabsize=2,breaklines]{haskell}
define c_1_0 =
     ((Self == D0) /\ False) \/
     (((Self == D1) /\ False) \/
     ((Self == D2) /\ False))

define c_1_1 =
     ((Self == D0) /\ False) \/
     (((Self == D1) /\ False) \/
     ((Self == D2) /\ False))

define c_2_0 =
     ((Self == D0) /\ ((time == 2) /\ values_received[0])) \/
     (((Self == D1) /\ ((time == 2) /\ values_received[0])) \/
     ((Self == D2) /\ ((time == 2) /\ values_received[0])))

define c_2_1 =
     ((Self == D0) /\ ((time == 2) /\ values_received[1])) \/
     (((Self == D1) /\ ((time == 2) /\ values_received[1])) \/
     ((Self == D2) /\ ((time == 2) /\ values_received[1])))
\end{minted}

We see that MCK has calculated that there is not common knowledge of either value at time 1,
and that at time 2, an agent has common knowledge that some agent $i$ had an initial value of $\mathtt{v}\in \{0,1\}$  iff
$\mathtt{values\_received[v]}$ holds. (This local variable is an alias for the variable $\mathtt{w[i][v]}$ in the environment.)
As a result, the formulas above are all evaluated to hold by the model checker for the
concrete program produced by the synthesis process.
\end{arxiv}

\end{document}